\documentclass[12pt,preprint]{aastex}

\def\hide#1{}

\begin{document}

\title{Peak Luminosities of the Hard States of \mbox{GX~339-4}: Implications for the Accretion Geometry, Disk Mass, and Black Hole Mass\\}

\author{Wenfei Yu\altaffilmark{1,2}, Frederick K. Lamb\altaffilmark{2}, Rob Fender\altaffilmark{3,4}, \& Michiel van der Klis\altaffilmark{4}}
\altaffiltext{1}{Shanghai Astronomical Observatory, 80 Nandan Road, Shanghai, 200030, China }
\altaffiltext{2}{Center for Theoretical Astrophysics and Department of Physics, University of Illinois at Urbana-Champaign, 1110 W. Green St., Urbana, IL 61801. E-mail: wenfei@uiuc.edu}
\altaffiltext{3}{School of Physics and Astronomy, University of Southampton, Highfield, Southampton, SO17 1BJ, U.K. E-mail: rpf@phys.soton.ac.uk}
\altaffiltext{4}{Astronomical Institute, ``Anton Pannekoek'', University of Amsterdam, Kruislaan 403, 1098 SJ Amsterdam, The Netherlands. E-mail: michiel@science.uva.nl}

\begin{abstract}
We have analyzed observations of the black hole transient \mbox{GX~339$-$4} made with the {\it Rossi} X-ray Timing Explorer (RXTE) and the Burst and Transient Source Experiment (BATSE) on board the Compton Gamma-ray Observatory (CGRO). We have found a nearly linear relation between the peak flux during the low/hard (LH) state that occurs at the beginning of an outburst and the time since the flux peak of the latest LH state identified in the previous outburst. Assuming that the rate at which mass accumulates in the accretion disk between these peaks is constant and that any mass that remains in the disk after an outburst has a negligible effect on the next outburst, this nearly linear relation suggests that the peak flux during the LH state that occurs at the beginning of an outburst is related to the mass in the disk, and thus that the entire disk is probably involved in powering these LH states. We have also found a positive correlation between the peak luminosities of the LH state in the three recent outbursts of \mbox{GX~339$-$4} and the peak luminosities of the following HS state. This correlation is similar to the correlations reported previously for \mbox{Aql~X$-$1}, \mbox{4U~1705$-$44}, and \mbox{XTE~J1550$-$564}, providing further support for the idea that the accretion flow that powers the LH state is related to the accretion flow that powers the following HS state. Although the luminosity of the LH-to-HS transition varies by up to an order of magnitude, the neutron stars and the black holes are distinguishable in the state transition luminosity. We discuss the implications for the mass determination of the compact stars in the ultraluminous X-ray sources (ULXs). 
\end{abstract}

\keywords{accretion, accretion disks --- black hole physics --- stars: individual (\mbox{GX 339$-$4})}

\section{Introduction}
The low/hard (LH) state and the high/soft (HS) state are two main spectral states that have been identified in neutron star and black hole X-ray binaries (see van der Klis 1995; McClintock \& Remillard 2006). When a black hole or neutron star soft X-ray transient (SXT) rises in luminosity during an outburst, it may go through a state transition between the LH state and the HS state (note: a LH state outburst may occur). On its way back to the quiescent state, the transient may go through a state transition from the HS state to the LH state. 

Spectral states are solely determined by mass accretion rate in various accretion models (e.g. Esin et al. 1997). However, two phenomena associated with the state transitions hint that mass accretion rate is not the only parameter in determining the spectral states. One is that the luminosity corresponding to the state transition is not fixed (Maccarone 2003; Yu et al. 2004, here after YKF04; M. Nowak et al. 2004, private communication; Zdziarski et al. 2004). The other is that a source tends to remain in the HS state to lower luminosity levels during a decay, and remain in the LH state to higher luminosity levels during a rise (e.g Miyamoto et al. 1995), which is usually called hysteresis of state transitions. 

The two phenomena can be explained if the property of the accretion flow, probably reflected in the mass of the disk, affects the luminosity of state transitions, as suggested in the study of the hard-to-soft state transitions in \mbox{Aql~X$-$1}, \mbox{4U~1705$-$44} and \mbox{XTE~J1550$-$564} (YKF04). The study of the state transitions in an individual source can avoid the influence of the uncertainties in the estimates of many source quantities, such as those of the compact object mass and source distance. Here we report our study of the observations of the three recent major outbursts of the black hole transient \mbox{GX 339$-$4}.  We have found that the luminosity corresponding to the hard-to-soft state transition, i.e., the peak luminosity of the LH state before the transition, is positively correlated with the peak luminosity of the following HS state. This is similar to those seen in \mbox{Aql~X$-$1}, \mbox{4U~1705$-$44} and \mbox{XTE~J1550$-$564} (YKF04). This strengthens the idea that the accretion flow that powers the LH state and the accretion flow that powers the HS state are related. We have also found from BATSE and RXTE observations made in the past 15 years that the peak flux of the LH state at the beginning of an outburst is nearly proportional to the waiting time since the previous outburst, indicating a link between the flow that powers the LH state and the mass in the accretion disk.

\section{Observations, data analysis and results}

Using the method we described in YKF04, we have analyzed the standard data products of the public RXTE observations of \mbox{GX 339$-$4}. In Figure~1, we plot the ASM daily-averaged light curve, the PCA (2--9 keV) light curve, the HEXTE (15--250 keV) light curve, and the ratio between the count rates of the HEXTE (15--250 keV) and the PCA (2--9 keV). The typical HEXTE/PCA ratios corresponding to the LH state and the HS state are shown as a dotted line and a dashed line, respectively. In the HEXTE light curve, we have marked the peaks prior to the hard-to-soft transitions as {\it A, B, {\rm and} C}, and at the ends of the soft-to-hard state transitions as {\it $A_{e}$, $B_{e}$, {\rm and} $C_{e}$}, representing the peak LH states before the hard-to-soft transition and after the soft-to-hard transition, respectively.  {\it B} and {\it C} correspond to the peaks followed by the hard-to-soft transition in the outburst of 2002--2003 and the outburst of 2004, while {\it A} approximately corresponds to the peak associated with the hard-to-soft state transition in the outburst of 1998 (Kong et al. 2002). 

Then we measure the peak luminosities of the HS states as well as those of the LH states from spectral analysis. We have selected the observation corresponding to the PCA (2--9 keV) count rate peak which represents the HS state luminosity peak in each outburst. The observation IDs are 30165-02-01-00, 60705-01-81-00, and 70110-01-33-00, respectively. These observations are consistent with those peaks shown in the ASM light curve and are about 100 days after the corresponding peaks of the LH states seen in the HEXTE light curve. We have made use of the 2--30 keV PCA data and the 20--200 keV HEXTE data to generate the energy spectra. We have fitted the corresponding energy spectra with a model composed of an absorption by neutral hydrogen with a column density of $6\times{10}^{21} ~{\rm atoms~ cm}^{-2}$, a disk black body component representing the emission from an accretion disk, a Guassian line cenered at 6.5 keV representing the iron liine, and a power-law component with no cut-off, which has been seen in the HS state of black hole binaries (e.g., Zhang et al. 1997). A systematic error of 1\% or less was included in the model fits. All spectral fits are acceptable with a reduced $\chi^2$ of 2 or less. Assuming a source distance of 8 kpc (Zdziarski et al. 2004), we have obtained the 2--200 keV peak luminosities of the HS states from the spectral fits. We have also estimated the bolometric luminosities by considering the bolometric luminosity of the black body component below 2 keV. The estimated bolometric luminosities for the HS state peaks are about twice the 2--200 keV luminosities in individual outburst, indicating that the estimate of the bolometric luminosity of the HS state is strongly dependent on the assumption in the soft X-ray band.

Similarly, we have derived the peak luminosities of the LH states corresponding to {\it A, B, {\rm and} C} (i.e., the luminosity corresponding to the hard-to-soft transition).  The corresponding observations are 20181-01-07-01, 60705-01-69-00, and 70110-01-08-00. The PCA data in the energy range between 2 and 30 keV and the HEXTE data in the energy range between 20 and 200 keV were used for this analysis. In the spectral fits, we have applied a spectral model consisting of an absorption by neutral hydrogen with a column density of $6.0\times{10}^{21} ~{\rm cm}^{-2}$, a disk blackbody component representing an accretion disk emission, a cut-off power-law component approximately representing the Comptonization process in the system, and a Gaussian line centered at 6.5 keV representing the iron line. 

\subsection{Correlation between the luminosity of the hard-to-soft transition and the peak  luminosity of the HS state}
In Figure~2, we plot the 2--200 keV peak luminosity of the HS state versus the 2--200 keV peak luminosity of the LH state of the three outbursts as triangles, in comparison to those of \mbox{Aql~X$-$1} (filled circles), \mbox{4U~1705$-$44} (diamonds), and \mbox{XTE~J1550$-$564} (squares), which were obtained from spectral fits with the same models to the data in the observations reported in YKF04 (note: for the two neutron star systems an additional single blackbody component was included and the black body components are not always required in the fits, see also Yu \& Dolence 2006). Notice again that the 2--200 keV peak luminosity of the LH state is the luminosity corresponding to the hard-to-soft state transition in these cases. The peaks of the two flares seen in the ASM light curve of \mbox{4U~1705$-$44} were not covered by RXTE pointed observations, so we use one observation of the HS state in its second flare to estimate the peak luminosities of the HS states for the two flares, by scaling the corresponding ASM (2--12 keV) rates.  The typical values of the column density $N_{H}$ of 6.0$\times{10}^{21}$ (\mbox{GX 339$-$4}: Zdziarski et al. 2004), 3.4$\times{10}^{21}$ (\mbox{Aql~X$-$1}: Maccarone \& Coppi 2003), 2.4$\times{10}^{22}$ (\mbox{4U~1705$-$44}: Olive et al. 2003) and $9.0\times{10}^{21} {\rm cm}^{-2}$ (\mbox{XTE~J1550$-$564}: Kaaret et al. 2003) were used in the spectral fits. We have also used the source distances often used in previous studies. They are  8.0, 4.0, 7.4 (Haberl \& Titarchuk 1995, measured from type-I X-ray burst), and 6.0 kpc for the four sources, respectively. There is a strong correlation between the peak luminosities in \mbox{GX 339$-$4} (see Fig.~2), similar to those found in the other sources. If we describe the relation in each source in a form of $\log({L_{\rm HS, 37}})=S\log({L_{\rm ST, 37}})+C$, where $L_{\rm HS, 37}$ and $L_{\rm ST,37}$ are the peak luminosity of the HS state and the luminosity corresponding to the hard-to-soft  state transition in units of $10^{37}$ ergs/s, and $C$ is a parameter (0.34 for \mbox{GX 339$-$4}), we find the slope $S$ is between 0.6 (\mbox{GX 339$-$4}) and 1.3 (\mbox{XTE~J1550$-$564}). Because black hole binaries have more powerful jets than neutron star binaries in the LH state (Fender \& Kuulkers 2001) and the jet is quenched in the HS state (Fender et al. 1999), the similar slopes and displacements in \mbox{GX 339$-$4} and \mbox{Aql~X$-$1}, which are independent of the distance assumed,  suggests that the jet contribution to the X-ray flux in the bright LH states should be very small. The estimated luminosity corresponding to the hard-to-soft state transition is affected by the source distance used. The measurements of the distances of \mbox{GX 339$-$4} ($\gtrsim$ 6.0 kpc, Hynes et al. 2004) and \mbox{XTE~J1550$-$564} ($\gtrsim$ 3.2 kpc, Orosz et al. 2002) are quite uncertain, which would place the lowest luminosity of the hard-to-soft state transition in \mbox{GX 339$-$4} and \mbox{XTE~J1550$-$564} above 2.0 and $3.0\times{10}^{37} {\rm ergs~s^{-1}}$, respectively. On the other hand, the distance to \mbox{Aql~X$-$1} (between 4 and 6.5 kpc; see Rutledge et al. 2001) would put its lowest value of the luminosity of the hard-to-soft state transition below $2.6\times{10}^{37}~{\rm ergs~s^{-1}}$, while a possible overestimate of the distance of \mbox{4U~1705$-$44} from type I X-ray bursts would put its lowest luminosity of state transition below $2.0\times{10}^{37}$ ergs/s. The boundaries of the neutron star systems and the black hole systems in Fig.~2 overlap in the narrow range $(2-3)\times{10}^{37}~{\rm ergs~s^{-1}}$, a result of large uncertainties in the estimates of the source distances. 

\subsection{Correlation between the peak flux of the initial LH state and its waiting time in the outbursts}
We have obtained the BATSE light curves of \mbox{GX 339$-$4} in three energy bands (20--40, 40--70, and 70--160 keV) from BATSE group website (http://www.batse.msfc.nasa.gov/batse/occultation/list.html). We have averaged these light curves to get a 10-day averaged light curve in the energy range between 20 and 160 keV. We have also extracted the RXTE/HEXTE daily averaged light curve of \mbox{GX 339$-$4}, and have converted the HEXTE (20--250 keV) count rates into fluxes in BATSE crab units. To do this, we have matched the light curves of the 1996--1998 outburst from both instruments, which gives that 305 counts $s^{-1}$ in the HEXTE (20--250 keV) light curve is equvalent to 1 crab in the BATSE light curve (20--160 keV).  The combined BATSE and HEXTE light curves are shown in Figure~3. 

There are a total of seven major outbursts seen in \mbox{GX 339$-$4} in the past 15 years. In Figure~3, we have marked the initial hard X-ray peaks of each outburst as 1--7, corresponding to the initial peaks of the LH states. The hard X-ray peaks at the end of the soft-to-hard transition in the decays of the outbursts 5--7 are marked as $5_{e}$, $6_{e}$ and $7_{e}$, respectively. Peak 1 was probably followed by a very short ($\sim 1-2$ day) HS state, indicated in the Ginga light curve showing a sharp rise and decay (Kong et al. 2002). For the peaks 1--4, the outburst intervals are short and the hard X-ray fluxes between the peaks are low compared with the peak fluxes of $5_{e}$--$7_{e}$, suggesting that the outbursts 1--4 correspond to either LH state outbursts, or correspond to outbursts associated with a very short HS state, whose duration is negligible compared to the corresponding peak-peak interval. The durations of the outbursts 1--4 are about 100--150 days. The intervals between the LH state peaks in the outburst rises (i.e., peak-peak intervals seen in BATSE outburts) for outbursts 1--7 are approximately equal to the sum of the corresponding outburst quiescent interval plus the duration of a rise of the LH state in an outburst and the duration of a decay of the LH state in the previous outburst. The outburst intervals of the last two peaks, namely 6 and 7, were determined from the intervals between $5_{e}$ and 6, and between $6_{e}$ and 7, respectively. For example, the time between the peak flux of the LH state at the beginning of the outburst 6 and the last flux peak of the LH state in the outburst 5 is the time interval between the two peaks $5_{e}$ and $6$, which includes the decay time of the LH state from the peak $5_{e}$, the quiescence time between outburst 5 and outburst 6, and the rise time of the LH state the peak $6$. Thus outburst intervals are actually the waiting times of the LH peaks. We have found a nearly linear relation between the peak flux of the initial LH state flare (peaks 2--7) and the waiting time, with a linear Pearson's correlation coefficient of 0.998. The nearly linear relation is shown in the inset panel of Fig.~3. It is worth noting that in each outburst the sum of the duration of a decay of the LH state to quiescence and the duration of the rise of the LH state from quiescence in the next outburst is approximately equal to the duration of a single LH state outburst, as seen in the outburst peaks 1--4 and 7, with a similar duration of about 100--150 days. Therefore a nearly linear relation of the same slope as the one shown in Fig.~3 can be inferred for the relation between the peak flux of the LH state and the quiescent time between the outbursts. The intersection between the linear relation and the time axis would give a negative value of 100--150 days. This suggests that the source can rise to $\sim$ 0.1 crab without actually returning to its quiescence. This is consistent with the observation of several flaring events in the LH state in the beginning of the fifth outburst in Fig.~3.

A linear fit to the empirical relation shown in Figure~3 gives the peak flux $F_{\rm p}=(9.3\pm0.1)\times {10}^{-4}~T_{\rm w} - (0.036\pm0.01)$ in units of crabs, as seen with the BATSE in the 20--160 keV energy band, where $F_{\rm p}$ is the peak flux and $T_{\rm w}$ is the waiting time in units of days. If written in units of the 15--250 keV HEXTE count rates (two clusters) based on the RXTE observations since 1996, the relation is $C_{\rm HEXTE}=0.28~T_{\rm w}-10.87$ counts $s^{-1}$. Using these, we can predict the peak flux of the LH state of the next outburst of \mbox{GX 339$-$4} when the beginning of the outburst is determined. For example, if we detect a rise of \mbox{GX 339$-$4} in 2006, we have $C_{\rm HEXTE} = 0.2824\times (256 + {\rm Day~of ~2006)} - 10.87$ counts $s^{-1}$, with a model uncertainty of 2.5 counts $s^{-1}$. We have noticed that around mid-March to early-April of 2006, \mbox{GX 339$-$4} was seen by ASM at a peak rate $\sim$ 3 counts $s^{-1}$. Taking the time of the peak as 70th day of 2006, the model gives a HEXTE count rate of 81 counts $s^{-1}$. There is an additional uncertainty of 8 counts $s^{-1}$, or 0.025 BATSE crabs, estimated from the scattering of the data of the previous outbursts from the linear relation seen in the inset panel of Fig.~3. The RXTE actually observed \mbox{GX 339$-$4} in this period with weekly pointed observations. The corresponding priority data can be used to test the empirical relation. 

\section{Discussion and conclusions}

\subsection{Relation between the peak flux of the LH state and the waiting time}
We have analyzed the observations of \mbox{GX 339$-$4} made using the BATSE and the RXTE/HEXTE over the past 15 years and have found that the peak flux of the LH state at the beginning of an outburst is nearly proportional to the time since the last flux peak of the LH state in the previous outburst (see subsection 2.2 and Fig.~3). Thus, if the rate at which mass accumulates in the accretion disk is constant and the mass remaining in the disk after the end of an outburst has a negligible effect on the peak flux of the LH state in the next outburst, then the peak flux of the LH state is proportional to the mass stored in the accretion disk. This apparent link between the peak flux of the LH state and the total mass accumulated in the disk suggests that the accretion flow that powers the LH state extends throughout the disk. It is worth noting that when Eddington limit is reached, there would be a saturation of the peak flux. 

With the above assumptions, the nearly linear relation obtained from the observations of the outbursts with and without a hard-to-soft state transition suggests that the peak flux of the LH state is proportional to the mass in the disk, regardless of whether a hard-to-soft state transition follows. Therefore  during the outbursts with no hard-to-soft state transition the source actually underwent failed hard-to-soft state transitions and the disk accretion flow could never dominate in the inner-most region. Miller et al. (2006a,b) have shown the evidence for an outflow in several black hole binaries. It is possible that a higher mass loss rate from the disk to the outflow or jet is associated with these failed hard-to-soft transitions, since the disk flow failed to dominate. 

Assuming the LH states are the same, our result suggests that the peak flux of the LH state after the soft-to-hard transition is also linearly related to the mass in the disk. If so, the peak flux of the LH state should be lower than the peak flux of the LH state before the hard-to-soft transition, since some mass in the disk should have been accreted during the HS state. This can be used to explain the hysteresis effect in state transitions (e.g., Miyamoto et al. 1995; and the observations reported in this paper). If this is true, we would expect a correlation between the total integrated luminosity of the HS state in an outburst and the flux drop between the peak flux of LH state in the rise and the peak flux of the LH state in the decay. This deserves future investigations. 

\subsection{Correlation between the peak Luminosities of the LH state and the HS state}
We have also analyzed the RXTE observations of the three recent major outbursts of \mbox{GX 339$-$4} and have found a positive correlation between the peak luminosity of the low/hard (LH) state (i.e., the luminosity of the hard-to-soft state transition in these cases) and the peak luminosity of the following high/soft (HS) state, similar to the correlations previously found in the outbursts of \mbox{Aql~X$-$1}, \mbox{4U~1705$-$44}, and XTE J1550$-$564 (YKF04). The result suggests that the accretion flow that powers the LH state at the beginning of an outburst is related to the accretion flow that powers the following HS state (YKF04). It is worth noting that a saturation of the correlation is expected when the Eddington limit is reached in the HS state as well as in the LH state. 

There are two possibilities for the occurrence of the correlation between the peak luminosities. One is that in addition to the correlation between the peak luminosity of the LH state and the mass stored in the disk, the peak luminosities of the HS states in the outbursts are proportional to the mass stored in the accretion disk as well. This leads to a correlation between the peak luminosities of the LH state and the following HS state. Consider an outburst that evolves in luminosity as $L(t)$. $L(t)$ is $0$ when time $t$ is outside the outburst time interval [0, $\Delta t$], where $\Delta t$ is the duration of the outburst. The maximum of $L(t)$ is $L_{\rm peak}$.  If all outbursts have similar profiles and durations, the instantaneous luminosity is thus proportional to $L(t)$, i.e., $kL(t)$, where $k$ is a parameter of each outburst. The peak luminosity of the outburst is then $kL_{\rm peak}$, and the total energy of the outburst is $\int^{\Delta t}_{0} kL_(t)dt$, both proportional to $k$. As the total energy of an outburst is the total gravitational potential energy release of the mass in the disk accreted before the outburst, the mass in the disk is then proportional to $k$. This indicates that the peak luminosity of the outburst appears to correlate with the mass stored in the disk in outbursts of similar profiles. Because similar outburst duration requires that the mass stored in the disk is accreted in roughly the same time interval, the radial velocity of the disk flow has to be similar among the outbursts with different disk masses if disk size is about the same. This suggests that disk surface density in such outbursts has to be proportional to $k$ too, which can lead to different properties of the accretion flow during different outbursts. 

The other possibility is that the accretion flow that powers the LH state and the accretion flow that powers the HS state are directly related. The time lag between the occurrences of the two peak luminosities is about 100 days for \mbox{GX 339$-$4}. This corresponds to the viscous timescale of a standard $\alpha$ disk flow beyond a radius $R\sim{10}^{5} R_{g}$ for a 10 solar mass black hole. This radius is an order of magnitude larger than the transition radius between the standard disk and the central hot flow in the advection-dominated accretion flow model (Narayan et al. 1996), indicating the outer disk contributes to the generation of the energy flux of the LH state. This picture is similar to the two-flow picture discussed by Smith et al. (2002), but is different in the sense that the disk flow is the primary flow and the mass accretion rate corresponding to the corona flow may be an integration of the mass accretion rate throughout the disk at a given time. 

These results suggest an accretion flow geometry during an outburst rise is composed of  two distinct but related flows, i.e., a coronal inflow or outflow that primarily powers the LH state and a disk flow that primarily powers a possible HS state at a later time. Given that the correlation between the luminosity corresponding to the hard-to-soft state transition and the peak luminosity of the HS state holds in several other sources, the relation between the mass in the disk and the peak flux of the LH state, and that the hard-to-soft state transition can only occur after the LH state reaches its peak flux, which is set proportional to the mass in the disk, should be common. 


However, \mbox{GX 339$-$4} probably has a stable mass accumulation rate in the disk (mass inflow rate $-$ mass outflow rate) and the mass remaining in the disk after an outburst has a negligible effect on the next outburst. We may not see such a linear relation between the peak flux of the LH state and the outburst waiting time in other sources. For example, the property of an outburst in Aql X-1 depend on the properties of the previous outbursts (e.g., Simon 2001), so that a linear relation between the outburst luminosity or fluence and the waiting time cannot be seen. Therefore, \mbox{GX 339$-$4} is probably unique. 

\subsection{State transition luminosity, disk mass and black hole mass}
The soft-to-hard state transition occurs in a narrow luminosity range in Eddington units in black hole systems as well as in neutron star systems (see Maccorone 2003 and references therein). This means that the luminosity threshold for the soft-to-hard state transition is primarily set by the mass of the compact object and the effect from the mass in the disk is negligible. In Fig.~2, the neutron star systems are in the bottom-left corner, while the black hole systems are at the top-right corner. The hard-to-soft state transition in neutron stars is lower than that in the black holes. Our results in Fig.~2 therefore support a rough relation between the luminosity of state transition and the mass of the compact object. If we can distinguish neutron stars and black holes in the Galactic X-ray binaries, we would distinguish intermediate-mass black holes and stellar-mass black holes in the ultra-luminous X-ray sources (ULXs) in nearby galaxies by observing state transitions between the LH state and the HS state. 

We have shown that the mass in the disk probably affects the state transition luminosity in \mbox{GX 339$-$4}. This idea can be used to explain the stable luminosity of the state transitions in Cyg X$-$1. As a high mass X-ray binary (HMXB), wind accretion is dominant and the accretion disk should be relatively small. A small disk accreting at a certain mass accretion rate close to the state transition threshold has less mass than a larger disk accreting at the same mass accretion rate.  This would lead to the fact that the luminosity of state transition in Cyg X$-$1 is close to the luminosity solely determined by the mass accretion rate and therefore nearly constant (e.g. Zhang et al. 1997), similar to those of the soft-to-hard state transitions (see e.g., Maccarone 2003). A previous proposal of a two flow geometry (Smith et al. 2002) explained the constant state transition luminosity in terms of the disk size being too small (see also Zdziaski et al. 2004). We think it is because the mass of the disk is small and therefore the effect of the variation of the mass is negligible. 

The empirical form or the relation of the state transition luminosity to the compact object mass and the disk mass is not yet known. If the mass of the compact object and the mass in the accretion disk comprise an additional form, the relation between the state transition luminosity and the compact object mass would be determined when the mass in the accretion disk is negligible. The observations of the soft-to-hard state transition at the end of an transient outburst and those of the HMXBs would be used to determine the empirical relation. 

\acknowledgments
We would like to thank the RXTE GOF for providing the standard products to the users and Mike Nowak of MIT for helpful discussions during an ASPEN workshop in 2005. WY appreciates Jean Swank of the RXTE GOF for her kind help in his research, including scheduling TOO observations related to the research topic. This work was supported by NASA grants NAG 5-12030, NAG 5-8740, and NNG 05GL60G,  NSF grant AST 0098399, and funds of the Fortner Endowed Chair at Illinois, and by the One Hundred Talents Project of the Chinese Academy of Sciences initiated at Shanghai Astronomical Observatory (SHAO). This work has made use of data obtained through the High Energy Astrophysics Science Archive Research Center Online Service, provided by the NASA/Goddard Space Flight Center. When we were submitting this paper, we noticed a report that {\it Swift} had detected \mbox{GX 339$-$4} at 230 mcrab on 2006 Dec. 18. This outburst may be used to test the empirical relation determined in this paper.

\clearpage

\begin{figure*}
\epsscale{0.90}
\plotone{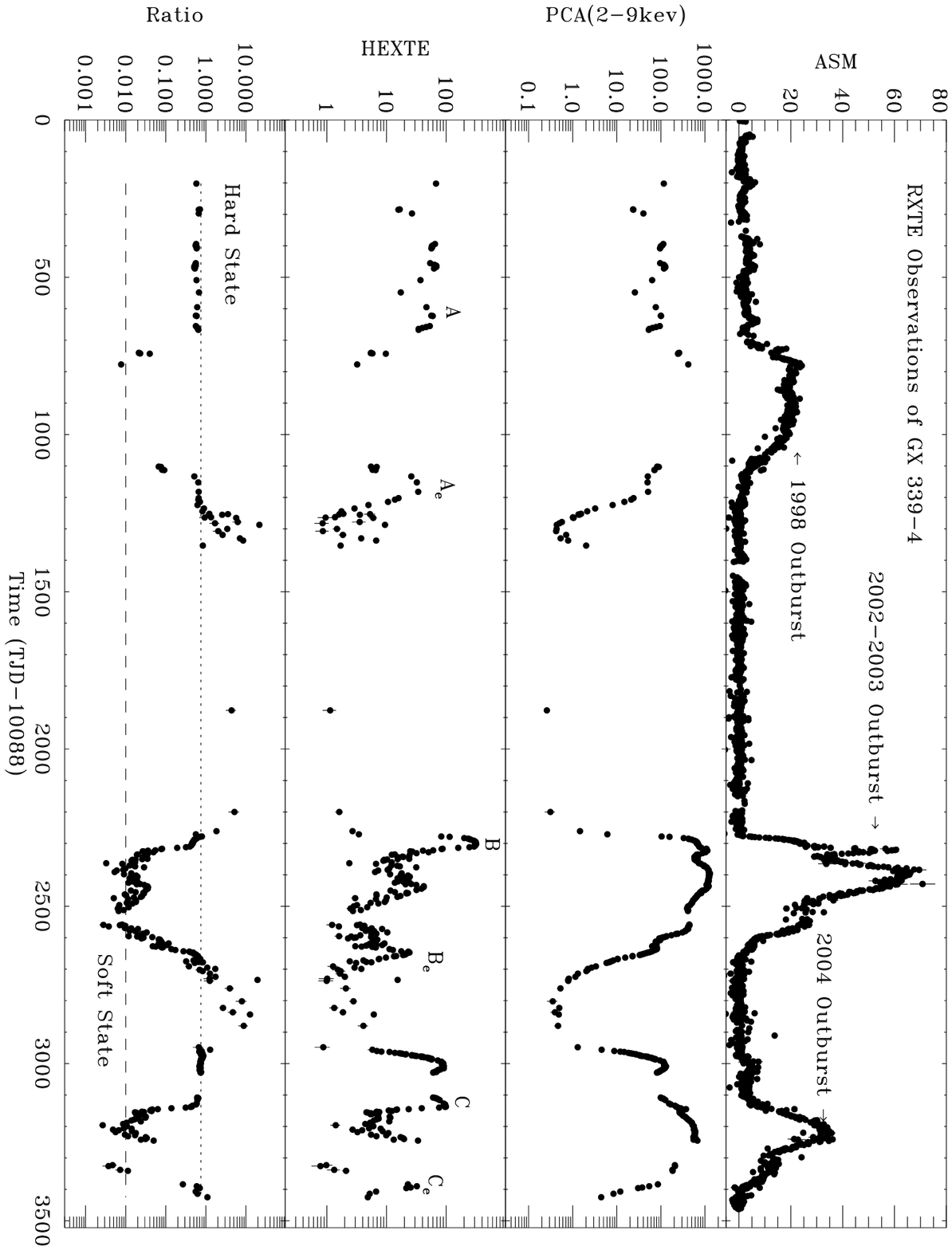}
\caption{The ASM (2--12 keV), the PCA (2--9 keV), the HEXTE (15--250 keV) light curves and the HEXTE(15--250 keV)/PCA(2--9 keV) count rate ratio of \mbox{GX 339$-$4} in the past nine years. The HEXTE peaks corresponding to the start of the hard-to-soft state transition and the end of the soft-to-hard state transition for the 1998 outburst are marked as {\it A {\rm and} $A_{e}$ }, for those of the 2002--2003 outburst are marked as {\it B {\rm and} $B_{e}$ }, and for those of the 2004 outburst are marked as {\it C {\rm and} $C_{e}$}. {\it A} is approximately at the same flux as the hard X-ray peak before the hard-to-soft state transition (see Kong et al. 2002).  Notice that the HEXTE peak rates corresponding to {\it A, B, C} correlate with the ASM peak rates of these outbursts, and the peaks in the HEXTE band lead those in the ASM band. }
\end{figure*}

\begin{figure*}
\epsscale{1.0}
\plotone{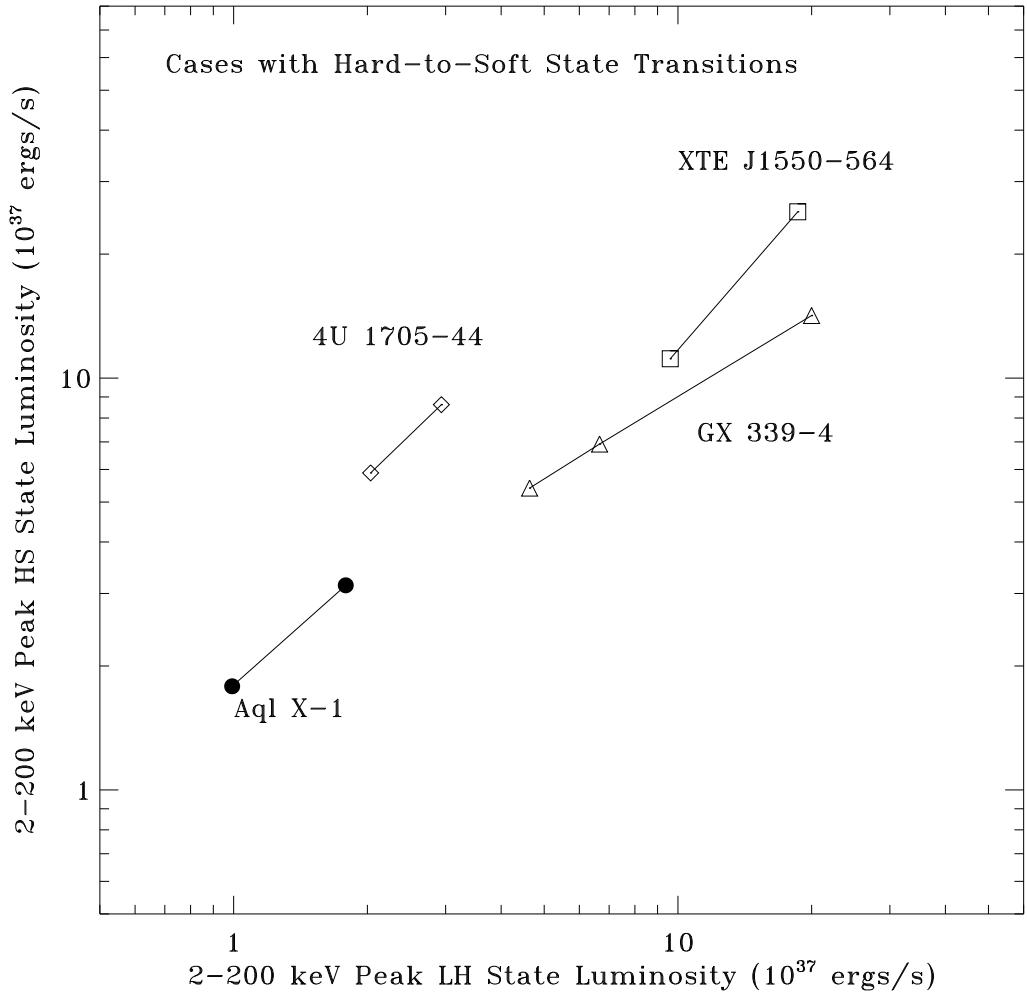}
\caption{The peak luminosities (2--200 keV) of the LH states and the subsequent HS states in \mbox{GX 339$-$4} (triangles) in the three outbursts. Observations of the sources studied in YFK2004 are also shown. Filled circles: \mbox{Aql~X$-$1}; Diamonds: \mbox{4U~1705$-$44}; Squares: \mbox{XTE~J1550$-$564}. Distances used to estimate these luminosities are 8, 4, 7.4, and 6.0 kpc, respectively. The relative uncertainties in the measured luminosities are around 3\% or less, which are not shown in the plot. A recent study of \mbox{Aql X$-$1} shows that the luminosity of the hard-to-soft state transition can be lower by a factor of 5 following the same linear relation (Yu \& Dolence 2006). }
\end{figure*}

\begin{figure*}
\epsscale{1.0}
\plotone{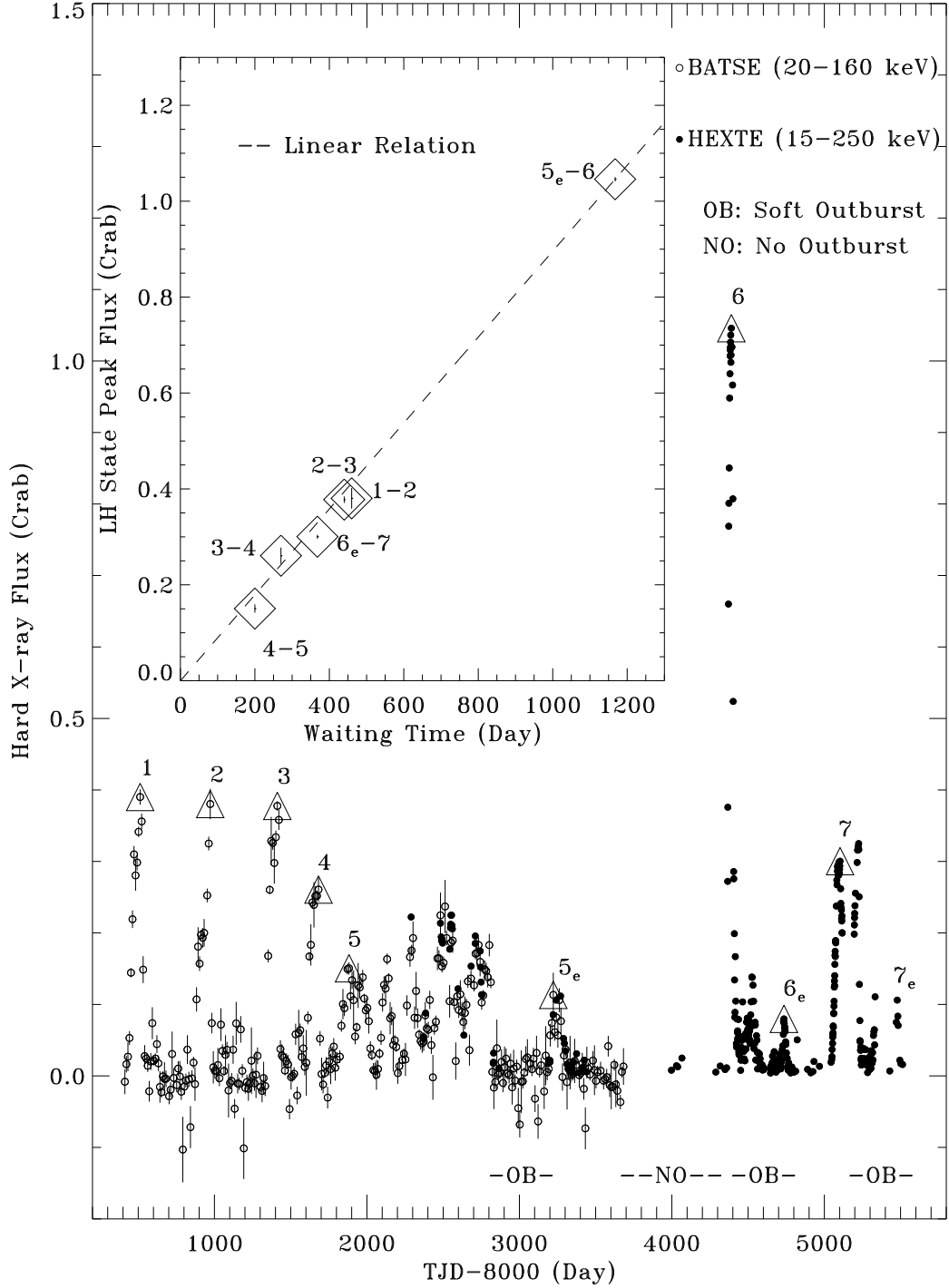}
\caption{The BATSE and HEXTE light curves of \mbox{GX 339$-$4} and a nearly linear correlation between the peak fluxes of the initial LH states in the outbursts of \mbox{GX 339$-$4} and the time since the latest LH state peak in the previous outburst (inset panel). The HEXTE (20--250 keV) count rates are divided by 305 so that the RXTE/HEXTE light curve match the BATSE light curve for the 1996--1998 outburst. The triangles indicate the LH state peaks used to determine the peak-to-peak waiting time. The diamonds are used only to call attention to the data points: their sizes have no meaning. The dashed line passes the origin and the data point of maximal peak flux, showing an example of a linear relation.}
\end{figure*}

\end{document}